\documentclass[preprints,review,accept,moreauthors,pdftex]{Definitions/mdpi} 
\pdfoutput=1

\usepackage{booktabs}

\DeclareMathOperator*{\argmax}{arg\,max} 

\newcommand{\bsh}{\texttt{BinarySkyHough}}
\newcommand{\crosscorrelation}{\texttt{Cross-Correlation}}
\newcommand{\powerflux}{\texttt{PowerFlux}}
\newcommand{\pyfstat}{\texttt{PyFstat}}
\newcommand{\eah}{\texttt{Einstein@Home}}
\newcommand{\falcon}{\texttt{Falcon}}
\newcommand{\fh}{\texttt{FrequencyHough}}
\newcommand{\gct}{\texttt{GCT}}
\newcommand{\sh}{\texttt{SkyHough}}
\newcommand{\soap}{\texttt{SOAP}}
\newcommand{\tdfs}{\texttt{Time-domain $\F$-statistic}}
\newcommand{\viterbi}{\texttt{Viterbi}}
\newcommand{\weave}{\texttt{Weave}}

\newcommand{\bsgl}{B_{\textrm{S}/\textrm{GL}}}
\newcommand{\bsgltl}{B_{\textrm{S}/\textrm{GLtL}}}
\newcommand{\bstar}{\mathcal{B}^{*}_{\textrm{S}/\textrm{G}}}
\newcommand{\dmoff}{\texttt{DM-off}}
\newcommand{\F}{\mathcal{F}}

\firstpage{1} 
\pubvolume{1}
\issuenum{1}
\articlenumber{0}
\pubyear{2021}
\copyrightyear{2020}
\datereceived{} 
\dateaccepted{} 
\datepublished{} 
\hreflink{https://doi.org/} 

\Title{Search methods for continuous gravitational-wave signals from unknown sources in the advanced-detector era}

\TitleCitation{Search methods for continuous gravitational-waves from unknown sources in the advanced-detector era}

\Author{Rodrigo Tenorio\orcidA{}$^{*}$, David Keitel\orcidB{}  and Alicia M. Sintes\orcidC{}}

\AuthorNames{Rodrigo Tenorio, David Keitel, Alicia M. Sintes}

\AuthorCitation{Tenorio, R.; Keitel, D.; Sintes, A.~M.}

\address[1]{Departament de Física, Institut d’Aplicacions Computacionals i de Codi Comunitari (IAC3),\\
Universitat de les Illes Balears, and Institut d’Estudis Espacials de Catalunya (IEEC),\\
Carretera de Valldemossa km 7.5, E-07122 Palma, Spain}

\corres{Correspondence: rodrigo.tenorio@ligo.org}

\abstract{
    Continuous gravitational waves are long-lasting forms of gravitational radiation produced by persistent 
    quadrupolar variations of matter. 
    Standard expected sources for ground-based interferometric detectors are neutron stars presenting 
    non-axisymmetries such as crustal deformations,  r-modes or free precession. 
    More exotic sources could include decaying ultralight boson clouds around spinning black holes. 
    A rich suite of data-analysis methods spanning a wide bracket of thresholds between sensitivity and
    computational efficiency has been developed during the last decades to search for these signals.
    In this work, we review the current state of searches for continuous gravitational waves using ground-based 
    interferometer data, focusing on searches for unknown sources. 
    These searches typically consist of a main stage followed by several post-processing steps to rule out 
    outliers produced by detector noise. 
    So far, no continuous gravitational wave signal has been confidently detected, 
    although tighter upper limits are placed as detectors and search methods are further developed.
}


\begin{document}


\section{Introduction}
\label{sec:introduction}
The search for continuous gravitational-wave signals (CWs), long-duration forms of gravitational radiation,
is one of the endeavours of gravitational-wave astronomy. 
These signals are produced by long-standing quadrupolar variations, 
such as rapidly-spinning neutron star (NSs) sustaining a crustal deformation,
undergoing an r-mode instability or in free precession~\cite{Lasky:2015uia, Glampedakis:2017nqy, 
Sieniawska:2019hmd, Haskell:2021ljd},
as well as more exotic sources such as the annihilation of ultra-light boson clouds around spinning 
black holes~\cite{Essig:2013lka, Brito:2017zvb, Brito:2017wnc, Zhu:2020tht}
or compact dark matter objects (CDOs) in the Solar System~\cite{Horowitz:2019pru}.

Detecting a CW signal could shed some light on NS physics, 
as well as open a new channel to test general relativity (extra polarizations, Lorentz violations)
or detect dark matter
\cite{Isi:2015cva, Isi:2017equ, Isi:2018pzk, Xu:2020zxs, Xu:2021dcw}.
No confident CW detection has been reported to date. 
The current product of CW searches are source-agnostic upper limits on the nominal CW amplitude $h_0$.
These results can then be mapped into different astrophysical scenarios,
such as the ellipticity of nearby NSs,
the mass of ultralight bosons around black holes~\cite{Palomba:2019vxe, Sun:2019mqb}, 
or the nearby population of planetary-mass
primordial black hole binaries~\cite{Miller:2020kmv, Miller:2021knj}.

The present document reviews search methods and pipelines employed to look for CW signals in the
observing runs performed by the second generation of ground-based interferometric detectors
(advanced detectors). 
Reviews on the physical mechanisms of CW emission by NSs can be found 
in~\cite{Lasky:2015uia, Glampedakis:2017nqy, Sieniawska:2019hmd, Haskell:2021ljd}.
Basic data analysis techniques are discussed in~\cite{Prix:2009oha}. 
Finally,~\cite{Riles:2017evm} discusses the main results of previous CW searches up to 2017.

The standard CW signal model consists of a quasi-monochromatic source emitting gravitational waves
at a certain frequency $f_0$. For the case of a NS sustaining a certain ellipticity, 
$f_0$ corresponds to twice the spinning frequency of the star. 
Further time derivatives of the frequency arise due to different physical mechanisms affecting
the source, such as energy emission as gravitational or electromagnetic radiation.
In the case of sources in binary systems, the orbital motion induces a Doppler modulation.

\begin{figure}
    \centering
    \includegraphics[width=0.49\textwidth]{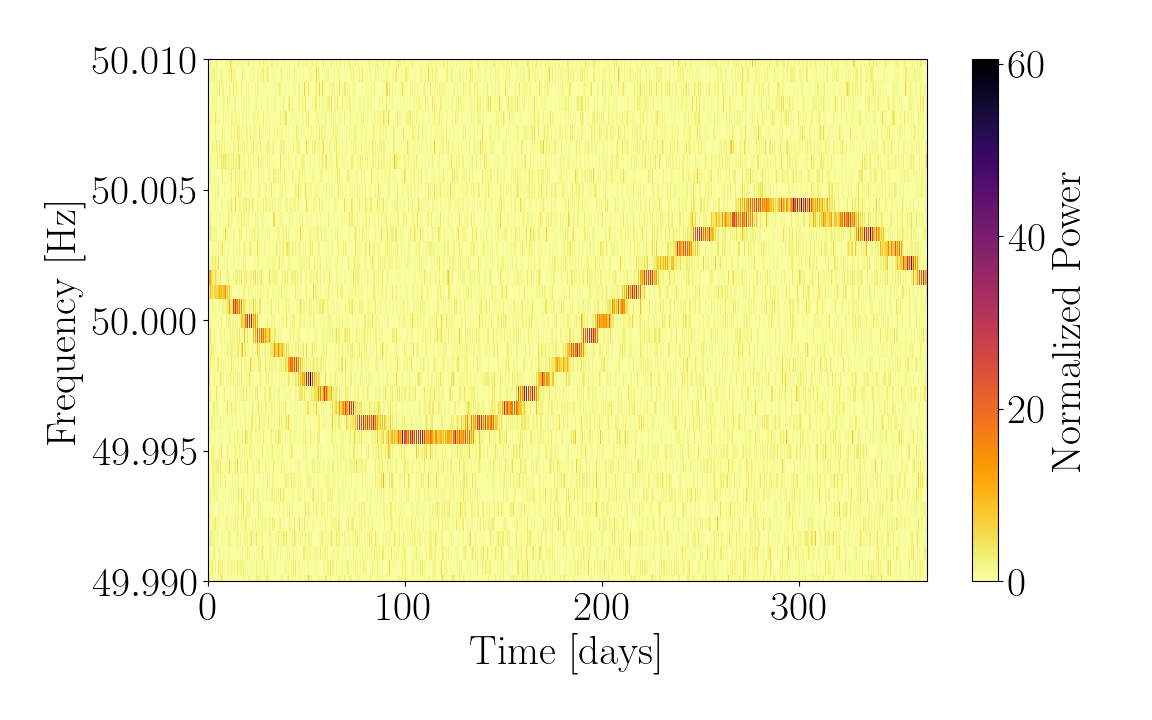}
    \includegraphics[width=0.49\textwidth]{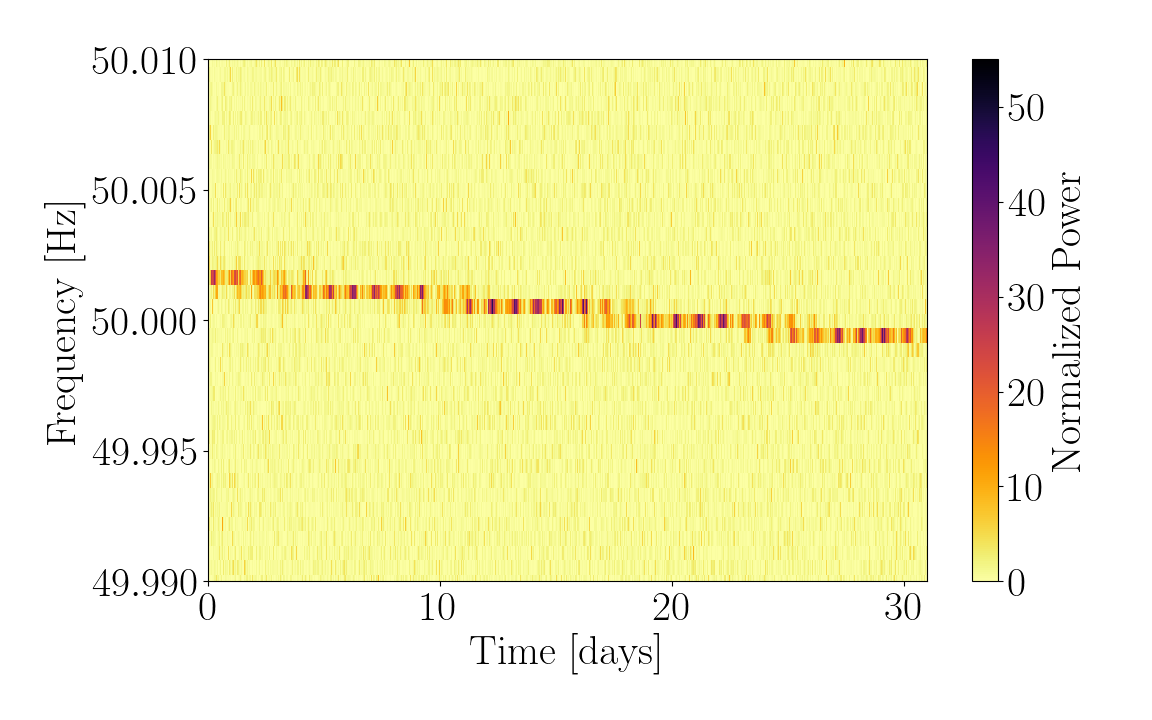}
    \caption{Data spectrogram showing the response of an Earth-bound detector 
    to a passing CW signal emitted by an isolated source.
    The left panel displays the one-year evolution of the detector response, 
    while the right panel zooms in on a month-long period. 
    Frequency modulations correspond to the yearly translation of Earth around the SSB 
    (the daily Doppler modulation is contained within a frequency bin). 
    Amplitude modulations correspond to the change in antenna-pattern functions throughout a day.}
    \label{fig:cw_spectrogram}
\end{figure}

Regardless of the specific intrinsic frequency modulation of the source, 
CW signals as seen from Earth are Doppler-modulated due to the detector motion
around the Solar System barycenter (SSB). 
For a source with a given intrinsic frequency evolution $\hat{f}$, 
the detector-frame frequency is given by
\begin{equation}
    f(t;\lambda) = \hat{f}(t) \left[ 1 + \frac{\vec{v}(t)}{c} \cdot \vec{n}\right]\;,
    \label{eq:track}
\end{equation}
where the phase-evolution parameters $\lambda$ include the CW frequency $f_0$ and the
sky position of the source $\vec{n}$, as well as any other parameter describing the intrinsic
frequency evolution of the signal. 
$\vec{v}(t)/c$ refers to the detector velocity expressed as a fraction of the speed of light, 
and contains both the daily and yearly motion of Earth around the SSB, 
of orders $\mathcal{O}(10^{-6})$ and $\mathcal{O}(10^{-4})$, respectively.
The yearly modulations can be clearly seen in the left panel of Fig.~\ref{fig:cw_spectrogram};
the daily modulation is contained within a frequency bin.

The amplitude of a CW signal is described using four parameters, namely the initial CW phase
$\phi_0$, the sperical angles describing the orientation of the source $\{\psi, \cos\iota\}$,
and the nominal gravitational wave amplitude $h_0$. The response of a ground-based detector
to a passing CW is better described in terms of the so called JKS representation~\cite{Jaranowski:1998qm}
\begin{equation}
    h(t; \lambda, \mathcal{A}) = \sum_{\mu=0}^{3} \mathcal{A}^{\mu} h_{\mu}(t; \lambda)\;,
\end{equation}
where the four time-independent $\mathcal{A}^{\mu}$ depend on the four amplitude
parameters $\mathcal{A} = \{\phi_0, \psi, \cos\iota, h_0\}$ and the antenna-pattern response of the detector
is contained in the four quadratures $h_{\mu}(t; \lambda)$, which only depend on the phase-evolution parameters.
The basic effect of the antenna-pattern response on a CW signal is a daily amplitude modulation, 
clearly visible in Fig.~\ref{fig:cw_spectrogram}.

As opposed to the short signals produced by compact binary coalescences (CBCs), 
with typical durations between a few minutes and less than a second for current detectors, 
CW signals are expected to last for years, 
spanning several observing runs of the current and future generation of ground-based interferometric detectors
\cite{LIGOScientific:2014pky, VIRGO:2014yos, KAGRA:2018plz,Maggiore:2019uih, Reitze:2019iox}.
This difference in duration is crucial in terms of detecting and estimating the parameters of a CW signal.

Modelled searches are usually performed using \emph{matched filtering}~\cite{LIGOScientific:2019hgc}, 
comparing the datastream to a set of templates in order to find a high correlation. 
Due to the typical duration of a CW signal (spanning the entire observing run),
the required number of templates to perform a blind search is prohibitively high even 
for current computing standards~\cite{Brady:1997ji, Prix:2007ks, Wette:2014tca}.

The standard strategy, in a broad sense, 
is to reduce the effective length of the datastream by performing matched filtering over shorter segments; 
the segment-wise results can then be combined into a final statistic.
These kind of schemes are usually referred to as \emph{semicoherent} searches~\cite{Prix:2009oha}:
the segment-wise analysis is typically referred to as \emph{coherent},
as it compares the phase evolution of a signal with the datastream throughout a coherence time $T_{\textrm{coh}}$.
The resulting coherent filters are then combined incoherently (ignoring phase information) 
into a final detection statistic. 
This incoherent combination allows to recover part of the sensitivity 
lost due to the split of the initial datastream;
the final sensitivity, however, is lower than that of a fully coherent search.
\begin{figure}
    \centering
    \includegraphics[width=0.5\textwidth]{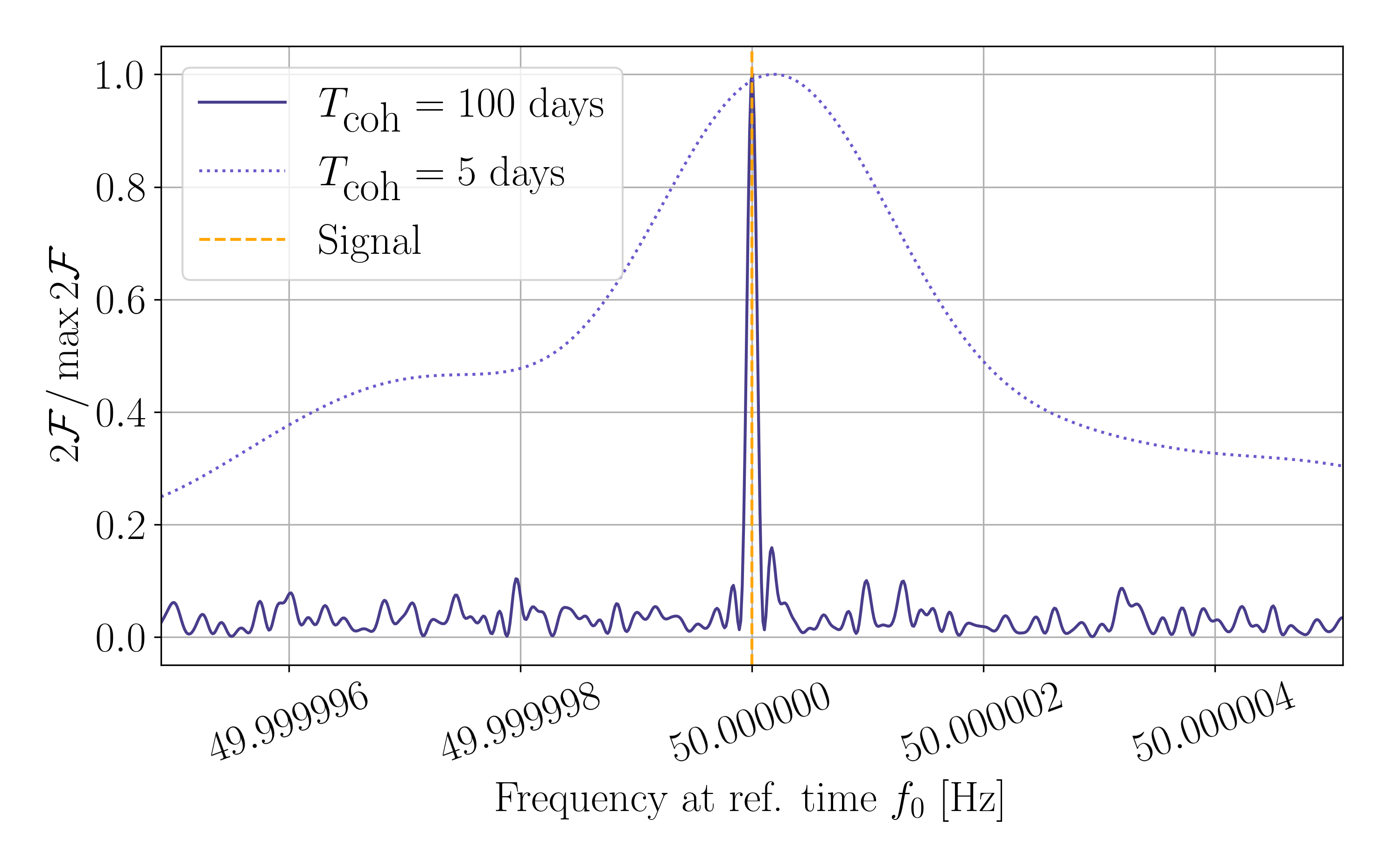}
    \caption{Effect of different values of $T_{\textrm{coh}}$ on one of the standard detection statistics for
    CW searches, the $\mathcal{F}$-statistic (see Sec.~\ref{sec:Fstat}). Longer coherence times impose a greater
    penalization to deviations with respect to the signal model; as a result, peaks tend to become narrower
    around the true signal parameters.}
    \label{fig:two_tcoh_grid_search}
\end{figure}
Figure~\ref{fig:two_tcoh_grid_search} illustrates the principle of operation of semicoherent searches.
By comparing shorter streams of data, 
a looser constraint is imposed when comparing phase-evolution templates. As a result, the characteristic
width of detection-statistic peaks widens, 
reducing the required number of templates to ensure a good covering of the parameter space.
This strategy is at the core of multi-stage approaches such as those discussed in Sec.~\ref{sec:follow_up}.

Parameter estimation, on the other hand, is positively affected by the long signal durations.
Typical frequency resolutions are under a mHz for initial stages, achieving nHz resolution
at fully-coherent follow-ups using a year of data. Sky localization is also significantly improved.
If we think of CBC sky localization, neglecting contributions from anntenna-pattern amplitude modulations and
higher GW modes, the problem is basically that of determining the direction a GW pulse came from,
which can be solved by means of measuring the pulse from $N$ detectors and finding the overlapping 
sky-positions~\cite{Fairhurst:2010is}. 
The case of CW signals is much simpler, as they are not just single pulses, 
but continuously arrive at the detector as it moves around the SSB.
Due to this movement, a \emph{single} interferometric detector recieving a CW signal at different positions 
with respect to the SSB is essentially equivalent to an arbitrary large set 
of \emph{different} detectors reciving the \emph{same} pulse from a source. 
Hence, for CW sources, sharp sky localization can be achieved \emph{using a single detector}
by simply extending the duration of an observing run.

CW searches require a very fine parameter-space resolution in order not to miss a signal,
increasing the computing cost of a matched-filtered search up to unaffordable figures~\cite{Brady:1997ji}. 
As a result, searches for CW signals from unknown sources tend to follow a hierarchical approach~\cite{
    Brady:1998nj, Cutler:2005pn, Prix:2012yu, Shaltev:2015saa, Papa:2016cwb,
    Ashton:2018ure,Tenorio:2021njf}: wide parameter-space regions are analized using a less-constraining 
statistic so that a coarser template bank can be used, see Fig.~\ref{fig:two_tcoh_grid_search}. 
Interesting regions are then typically small enough to follow up using a more sensitive statistic. 
A quantitative description of this strategy can be found in~\cite{Dergachev:2010tm}.

The structure of this work goes as follow: in Sec.~\ref{sec:searches} we review the main search methods
and pipelines employed to search for CW signals from unknown sources during the era of the advanced detectors.
Section~\ref{sec:post_processing} discusses different post-processing stages employed by said searches;
these include both specific prescriptions to select interesting candidates (e.g. clusterings) 
and consistency arguments to assess the overlap of a specific set of templates with instrumental disturbances.
Section~\ref{sec:follow_up} reviews follow-up strategies to further analyze interesting candiadtes. 
So far, no search has claimed a confident CW detection, 
reporting instead constraints on the maximum detectable amplitude achieved. 
Different approaches to construct said constraints are listed in Sec.~\ref{sec:sensitivity}.
Finally, a summary of the reviewed methods is presented in Sec.~\ref{sec:summary}.

\section{Wide parameter-space search pipelines}
\label{sec:searches}
We present a review of methods and pipelines employed to date to search for CW signals from unknown sources
during the era of the advanced detectors.
Specifically, we consider three kinds of searches for unknown sources:
(i) blind searches, with weak prior assumptions about possible source parameters~\cite{EaT_O1,
ASI_O1, ASI_O1_full, Dergachev:2019wqa, Dergachev:2019oyu, ASI_O2, Dergachev:2020fli, 
Dergachev:2020upb, Dergachev:2021ujz, Steltner:2020hfd, Covas:2020nwy, Wette:2021tbv, ASB_O3a, Tenorio:2021sfj, 
ASI_O3a};
(ii) spot-light searches, focused at sky regions harbouring an interesing population of objects
whose exact frequency is unknown, such as globular clusters or the Galactic Center (GC)~\cite{Dergachev:2019pgs, 
Piccinni:2019zub};
and (iii) directed searches, targeting specific celestial objects compatible
with a CW source, such as supernova remnants (SNRs) 
or low-mass X-ray binaries (LMXBs)~\cite{LIGOScientific:2017fer,
LIGOScientific:2017zez, LIGOScientific:2018esg, LIGOScientific:2019lyx, Ming:2019xse, Papa:2020vfz,  
Millhouse:2020jlt, Middleton:2020skz, Jones:2020htx, Lindblom:2020rug, Zhang:2020rph,
LIGOScientific:2021mwx,Ming:2021xtz}.

\subsection{$\F$-statistic searches}
\label{sec:Fstat}

The $\F$-statistic is a standard detection statistic for CW signals. 
Initially derived as a maximum-likelihood estimator with respect to 
amplitude parameters~\cite{Jaranowski:1998qm, Cutler:2005hc},
it was later rederived in a Bayesian context as a Bayes factor, 
gauging the presence (or lack) of a signal in a Gaussian noise data stream,
in which amplitude parameters are marginalized using a rather \emph{unphysical} 
set of amplitude priors~\cite{Prix:2009tq, Whelan:2013xka, Prix:2011qv, Dhurandhar:2017rlr, 
Bero:2018xyq, Wette:2021dmg}. 
This detection statistic can be extended for more generic types of sources,
such as binary white-dwarf systems~\cite{Prix:2007zh} 
or the inspiral phase of binary black-hole coalescenses~\cite{Keppel:2012ye}.

Semicoherent searches balance sensitivity and computing cost by choosing a suitable 
number of coherent segments whose combination into a semicoherent quantity can be performed in an efficient manner.
The basic ``stack-slide'' procedure used for many $\F$-statistic semicoherent searches~\cite{Stuart, Brady:1998nj, 
Cutler:2005pn, Prix:2012yu} is to set up a template bank in each coherent segment to compute
the segment-wise coherent detection statistic.
Then, in the semicoherent stage, 
the semicoherent detection statistic is computed on a \emph{finer} template bank
by combining results from the segment-wise coherent detection statistics.
This template bank refinement can be seen as a consequence of using multiple coherent segments:
the higher the number of coherent templates to be combined, 
the higher the resulting number of distinct semicoherent segments.
Discusion on the specifics of template bank refinements are deferred to 
Secs.~\ref{subsec:gct} and \ref{subsec:weave}.

We discuss three different implementations of semicoherent $\F$-statistic searches,
namely the Global Correlation Transform hierarchical search (\gct{}), \weave{} and \tdfs{}. 
They mainly diverge in the manner of constructing semicoherent quantities,
taking different trade-offs in terms of computing cost, memory requirement and robustness to non-Gaussianities. 
We note, however, that fully-coherent searches for targets at a specific sky position,
such as supernova remnants~\cite{LIGOScientific:2018esg, Lindblom:2020rug}, are still performed nowadays.

\subsubsection{\gct{} hierarchical search}
\label{subsec:gct}

Detection statistics (and the $\F$-statistic in particular) present a set of characteristic correlations
across the CW parameter-space due to some level of degeneracy present in CW signals~\cite{Prix:2005gx}. 
Understanding the structure of said correlations offers a simple method to reuse coherent $\F$-statistic 
values to construct a semicoherent statistic, reducing the overall computing cost of the search.
The \gct{} introduces a set of coordinates defined by the intersection of parameter-space correlation surfaces 
to identify nearby parameter-space points where the $\F$-statistic achieves a high value due to the presence
of a signal~\cite{Pletsch:2008gc, Pletsch:2009uu, Pletsch:2010xb}. 

Due to its interesting trade-off between sensitivity and computational cost~\cite{Walsh:2019nmr}, 
semicoherent \gct{} searches have been used by the \eah{} project
to perform deep searches throughout different observing runs~\cite{EaT_O1, Steltner:2020hfd, 
Ming:2019xse, Papa:2020vfz, Ming:2021xtz}. \eah{} searches distribute the computational load of a search
across a volunteer-computing network using BOINC~\cite{2019arXiv190301699A} to analyze a high number 
of parameter-space candidates.

At the core of the \gct{} search there are combinations of $\F$-statistic values computed at different 
coherent segments containing data from one or multiple detectors.
These are combined into the semicoherent $\F$-statistic using a finer template bank 
with refinement only on the spindown parameters~\cite{Pletsch:2009uu}.
This allows for the easy implementation of Bayesian extensions over the semicoherent $\F$-statistic,
such as the ``line-robust'' $\bsgl{}$ or $\bsgltl{}$ statistics~\cite{Keitel:2013wga, Keitel:2014gha, 
Keitel:2015ova, Keitel:2016wtx},
which combine $\F$-statistics from different detectors to suppress single-detector artifacts.

No general computing model is available for the \gct{} search,
requiring extensive software-injection campaigns in order to numerically tune the sensitivity of 
a search to the available computing resources~\cite{Papa:2016cwb}. 
As discussed in Sec.~IV B of~\cite{Wette:2015lfa}, this is due to a core assumption 
in~\cite{Brady:1998nj, Pletsch:2009uu, Pletsch:2010xb} neglecting refinements in the sky parameter-space.
Further developments in the field, discussed in Sec.~\ref{subsec:weave}, 
proposed a new strategy to solve this problem.
For the case of directed searches, however, optimal setup strategies are available~\cite{Ming:2015jla}.

\subsubsection{\weave{}}
\label{subsec:weave}

Local parameter-space structure can be understood in terms of the \emph{mismatch} 
$\mu$~\cite{Owen:1995tm, Prix:2006wm}.
Given a signal parameterized by a set of parameters $\lambda_{0}$, the parameter-space mismatch $\mu$ quantifies
the fractional loss of (squared) signal-to-noise ratio $\rho^2$ produced by an arbitrary parameter-space 
off-set $\Delta \lambda$ with respect to the true signal parameters:
\begin{equation}
    \mu(\Delta \lambda;\lambda_{0}) = 1 - \frac{\rho^2(\lambda_{0} + \Delta \lambda)}{\rho^2(\lambda_{0})} \;.
\end{equation}
In a neighbourhood of $\lambda_0$, the mismatch can be expressed in terms of a quadratic form as
\begin{equation}
    \mu(\Delta \lambda;\lambda_{0}) \simeq \Delta \lambda \cdot \mathbf{g} \cdot  \Delta \lambda \;,
\end{equation}
where the symmetric 2-rank tensor $\mathbf{g}$ plays the role of a Riemannian metric in the parameter-space.

It was quickly realized that, for the case of flat parameter spaces,
the metric $\mathbf{g}$ offers a complete description of the parameter-space, allowing for
the easy construction of optimal setups~\cite{Prix:2007ks, Allen:2021yuy}; alas, 
the standard CW parameter-space (specifically, the sky-position subspace) presents a non-trivial structure 
resulting in a curved parameter-space. 

\weave{}~\cite{Wette:2015lfa} represents the coming together of a number of search strategies.
It implements, for the first time, a semicoherent search with a well-understood computational model 
based on a flat parameter-space metric.
The setup is capable of constructing template banks at the suitable resolution
to achieve an optimal computing cost~\cite{Wette:2013wza, Wette:2014tca, Wette:2015lfa,
Wette:2016raf}.
Setting up \weave{} requires only a list of time stamps delimiting semicoherent segments, 
a coherent mismatch $\tilde{\mu}$, with which single-segment template banks will be constructed,
and a semicoherent mismatch $\hat{\mu}$, to setup the semicoherent template bank.
The identification of a template from the semicoherent bank to its corresponding ensemble
of coherent templates is handled by the \weave{} code using the coherent parameter-space 
metric to identify the nearest neighbor in each segment.
In a sense, \weave{} retains the general characteristics of the \gct{} search, 
mainly being an engine to combine coherent $\F$-statistics, but, as opposed to it, 
\weave{} requires no extensive numerical calibration to be deployed using an optimal setup.

The success of \weave{} as an all-sky search, however, is related to the characteristics of the
$\F$-statistic's parameter-space structure. 
As discussed in depth in~\cite{Wette:2016raf, Wette:2018bhc}, 
the optimal setup for a realistic computing budget 
tends to yield semicoherent mismatch values well beyond 
the validity of the metric approximation (i.~e.~$\mu \gg 1$). 
Nevertheless, empirical studies~\cite{Wette:2016raf} show that 
in this regime the $\F$-statistic actually falls off more slowly than predicted by the metric, 
meaning the resulting template bank will contain more templates than strictly required.
An alternative approach circumventing the empirical characterization of the $\mu \gg 1$ regime
is discussed in~\cite{Allen:2019vcl, Allen:2021yuy, Allen:2021eju}.

Despite achieving a better sensitivity than the \gct{} search at a fixed computing cost, 
the increased memory requirements of \weave{} make it so far unsuitable for 
its deployment on \eah{}~\cite{Walsh:2019nmr}.
Nonetheless, its sensitivity and setup flexibility have already been proved in the implementation 
of a novel CW search strategy~\cite{Wette:2021tbv}.

\subsubsection{\tdfs{}}

Both \gct{} and \weave{} searches rely on a common implementation of the $\F$-statistic~\cite{CFSv2}, 
publicly available under \texttt{LALSuite}~\cite{lalsuite}. 
The \tdfs{} pipeline~\cite{LIGOScientific:2014yew} uses a different implementation, 
based on the $\F$-statistic's time-dependent behavior throughout the observing run.
Instead of computing semicoherent quantities, it focuses on significant parameter-space points 
at coherent-segment level, looking for coincident candidates across different time segments and detectors.
Using this coincidence criterion (which we review in more depth in Sec.~\ref{sec:post_processing})
the pipeline automatically becomes robust to strong instrumental features,
since they tend to overlap with different parameter-space regions as an observing run progresses.

This particular implementation of the $\F$-statistic uses its own template bank setup in order to optimize
the number of fast Fourier transform (FFT) computations~\cite{Pisarski:2013dj, Poghosyan:2014mta},
which normally takes a significant part of the overall computing cost of the search.
Further improvements at parameter-estimation level are optimization algorithms to resolve the 
characteristic frequency multi-modality of the $\F$-statistic~\cite{2019CQGra..36v5008S} and the
inclusion of machine-learning algorithms to filter out non-astrophysical candidates~\cite{2019arXiv190706917M}.

\subsection{Fourier-transform-based searches}
\label{sec:sftsearches}

The efficiency of $\F$-statistic searches stems from the marginalization with respect to amplitude parameters, 
removing four parameter-space dimensions from the search~\cite{Prix:2007ks}.
Semicoherent pipelines, moreover, assume amplitude parameters to be independent across different coherent segments.
This approach makes it difficult to search for amplitude parameters, such as specific CW polarizations.

An alternative family of methods use Fourier transforms of short data segments (Short Fourier Transforms, SFTs)
as the basic unit of operation. 
The duration of an SFT is typically such that CW signals are contained within a single frequency bin.
For an isolated NS, this length is about 30 minutes~\cite{Krishnan:2004sv}, 
although the exact value depends on the considered frequency range.
We note that SFTs themselves can, in some situations, corresponds to coherent segments of a semicoherent search;
in such cases, the effective coherent length is proportional to the SFT length.

In order to construct a detection statistic $\mathcal{S}$, 
Fourier transform amplitudes $\tilde{x}$ are combined using a set of weights (effectively a kernel) 
$\mathcal{K}$ taking into account source polarization, 
antenna-pattern amplitude modulations, 
Doppler modulations and relative phase deviations across different detectors:
\begin{equation}
    \mathcal{S} = \sum_{t, t'} \tilde{x}^{*}(t) \tilde{x}(t') \mathcal{K}(t, t')\;,
    \label{eq:cross_corr}
\end{equation}
where parameters other than time dependency have been kept implicit for the sake of simplicity.
Frequency modulations are assumed to be contained within the specific set of Fourier amplitudes being 
combined, although $\mathcal{K}$ can be configured to increase robustness against different kinds of 
spectral leakage by including neighboring frequency bins into the kernel~\cite{Allen:2002bp,Dergachev:2010tm}. 

We review two families of searches stemming from Eq.~\eqref{eq:cross_corr}, depending on whether their primary
target is to increase the robustness against deviations from the intended CW model (\powerflux{} \& \falcon{})
or to improve sensitivity by increasing the effective amount of data used (\crosscorrelation{}).

\subsubsection{\powerflux{} \& \falcon{}}
\label{sec:falcon}

The \powerflux{} pipeline~\cite{PF, PF2,PFPolarization, Mendell:2007ww} estimates power from a CW source
depending on its sky position. The basic implementation~\cite{ASI_O1, ASI_O1_full, ASI_O3a}
uses a diagonal kernel $\mathcal{K}$ to combine Fourier power from each SFT.
In this case, the role of $\mathcal{K}$ is to diminish the contribution of unfavourable frequency bins 
(due to high noise floors or a low antenna-pattern response) to the total weighted Fourier power.
Since the final statistic ignores relative phase shifts between SFTs,
this corresponds to a semicoherent search for a specific CW polarization.

Loosely coherent methods~\cite{Dergachev:2010tm, Dergachev:2011pd, Dergachev:2018ftg} exploit the
flexibility of Eq.~\eqref{eq:cross_corr} to set up a kernel $\mathcal{K}$ to account 
for unmodeled phase shifts,
be it due to parameter-space mismatches or unaccounted physics 
such as small binary orbital modulations~\cite{Singh:2019han}.
To do so, phases are allowed to drift at most by a specific amount $\delta$ across contiguous data segments,
obtaining as a result~\cite{Dergachev:2010tm}
\begin{equation}
    \mathcal{K}(t, t'; \delta) = \left(\frac{\sin{\delta}}{\delta}\right)^{\left|t - t'\right|}\;,
\end{equation}
where data is assumed to come from a single detector for the sake of simplicity.
(The framework presented in~\cite{Dergachev:2010tm} is flexible enough to treat data from multiple detectors.)
This simple kernel illustrates the principle of operation of loosely coherent searches: 
if $\delta = \pi$, $\mathcal{K}$ behaves like a delta function and phases among contiguous segments 
are uncorrelated, effectively performing a semicoherent search;
if $\delta = 0$, $\mathcal{K}=1$ and phases are correlated throughout the full data stream,
performing a fully-coherent search.
Tuning $\delta$ to intermediate value allows to trade sensitivity and computing cost: 
low $\delta$ values involve simpler kernels (less non-zero terms), 
easing the computing cost of Eq.~\eqref{eq:cross_corr} but imposing tighter constraints
with respect to the chosen CW model. 
A discussion on physically relevant $\delta$ values can be found in~\cite{Dergachev:2010tm},
whereas an efficient implementation of low-$\delta$ kernels can be found in~\cite{Dergachev:2011pd, 
Dergachev:2018ftg}.

Due to its high computing cost, 
the initial implementation of loosely coherent methods on \powerflux{} was mainly used on directed
searches, such as spot-light surveys~\cite{Dergachev:2019pgs} or follow-up stages
(see Sec.~\ref{sec:follow_up}).
An efficient implementation, \falcon{}, was later developed to perform all-sky searches throughout
O1~\cite{Dergachev:2019wqa, Dergachev:2019oyu} and O2~\cite{Dergachev:2020fli, Dergachev:2020upb, 
Dergachev:2021ujz}, setting competitive constraints on the nearby population of
Galactic neutron stars.

\subsubsection{\crosscorrelation}
The \crosscorrelation{} search~\cite{Dhurandhar:2007vb, Whelan:2015dha, Wagner:2021hgv},
implemented in the LALSuite library~\cite{lalsuite},
closes the sensitivity gap between semicoherent and coherent searches by increasing the effective amount of data
using the correlation of Fourier amplitudes at different times.

Although initially designed to look for stochastic GW backgrounds, CW signals are a perfect target to be
looked for using \crosscorrelation{}, as they are long-lasting and deterministic.
This allows to use not only the cross-correlation of different instruments during a certain period of time,
but also the cross-correlation of data segments at different periods of time.
In the context of CW searches, \crosscorrelation{} searches have been employed to look for specific 
high-priority targets, 
such as CW emission from the LMXB system Scorpius X-1~\cite{LIGOScientific:2017zez, Zhang:2020rph}.
Complementarily, searches for stochastic GW backgrounds,
such as the \texttt{Radiometer} search~\cite{Mitra:2007mc, Ain:2015lea, Ain:2018zvo},
are able also to constrain CW amplitudes from specific sky locations, such as those of SNRs, 
LMXBs such as Scorpius X-1 or the 
Galactic Center~\cite{LIGOScientific:2016nwa, LIGOScientific:2019gaw, LIGOScientific:2021oez}.

To do so, the kernel $\mathcal{K}$ is constructed following the expected correlation of a signal across
different stretches of data at different times and detectors. Specifically, 
only amplitudes within a certain time range $T_{\textrm{max}}$ are combined together. 
As opposed to \powerflux{} and \falcon{}, however, polarization angles are averaged out using
uniform priors.
The $T_{\textrm{max}}$ parameter plays a similar role to that of $\delta$ (see Sec.~\ref{sec:falcon})
in terms of trading computing cost and sensitivity:
longer $T_{\textrm{max}}$ increases the number of cross-correlations to perform, 
but also imposes a tighter constraint with respect to the specified CW model.
Latest developments on this pipeline include the use of re-sampling techniques to accelerate its
evaluation~\cite{2018PhRvD..97d4017M} and the use of shear parameter-space transform to optimize
the setup of template banks~\cite{Wagner:2021hgv}.

\subsection{Hough-transform semicoherent searches}

Loud instrumental features in the data tend to saturate detection statistics due to their
strong resemblance to CW signals along short periods of time. This problem pushed forward the
development of methods capable of suppressing narrow-band features in the data, such
as the application of the Hough transform to the search for CW signals.

The basic idea is to limit the contribution of each semicoherent segment to a bounded quantity,
which contributes by a limited amount to the overall statistic. 
This is done by binarizing a power-like quantity (whitened Fourier power formatted as SFTs, 
as discussed in Sec.~\ref{sec:sftsearches}, or the $\F$-statistic) into
ones and zeroes using a predetermined threshold~\cite{Krishnan:2004sv}. 
Such a binarization, however, comes at the cost of ignoring noise-floor variations and antenna-pattern amplitude
modulations, meaning highly-contaminated frequency bands around low-sensitivity sky positions will contribute
the same amount as clean frequency bands at the most favored sky positions. 
This problem is usually solved introducing a set of weights into the detection statistic~\cite{Palomba:2005fp, 
skyhough_weights}.

The Hough transform~\cite{osti_4746348} can be used to identify shapes in binarized images.
Given a parameter-space parametrizing a family of curves, each set of parameters is assigned a score,
the number count, proportional to the number of pixels in the image consistent with the corresponding curve.
In CW searches, the parameter space is typically the set of phase-evolution parameters, according to which
the data spectrogram is traversed adding (weighted) ones and zeroes depending on whether each frequency bin
contains excess power or not.

Hough-transform based searches have been widely used during the latest all-sky 
searches~\cite{ASI_O1, ASI_O1_full, ASI_O2, Covas:2020nwy, ASB_O3a, Tenorio:2021sfj}.
We discuss the two main implementations of the Hough transform for CW searches, \sh{} and \fh{}.

\subsubsection{\sh{}}

The \sh{} pipeline was the initial implementation of the Hough transform to the search for CW signals
\cite{Krishnan:2004sv, Sintes:2006uc}. 
Due to the Earth's movement around the SSB, a CW signal arriving at the detector at a certain time $t$ 
with a given frequency $f_0$ has a very specific set of sky positions from which it could have originated.
Said sky positions take the shape of thick annuli
(``circles in the sky'')~\cite{Krishnan:2004sv, Prix:2005gx, Tenorio:2020cqm} which, for consistency, 
can be described themselves as one/zero regions in the sky patch. 
The binarization of the data spectrogram, thus, gets mapped into the selection and summation of 
sky patches containing different selected annuli. 

The ``circles in the sky'' are weakly affected by local changes in $f_0$; hence, once computed for a specific
frequency, these structures can be saved into a look-up table (LUT) to analyze several frequency bins.
The use of LUTs, on the other hand, affects the maximum sensitivity of \sh{} since only an approximated
frequency-evolution track is being used.
Further developments on the pipeline include the re-analysis of interesting candidates using the exact
frequency-evolution track and using more sensitive statistics in order to improve parameter 
estimation~\cite{SanchodelaJordana:2010kkv, PepThesis}. 
Also, the combination of LUTs can be easily accelerated using GPU parallelization~\cite{Covas:2019jqa}.

The use of LUTs depends on a basic set of CW parameters, namely frequency and sky position, meaning it
can be arbitrarily extended to look for different types of CW signals as long as the source's intrinsic and
extrinsic frequency evolutions are uncoupled from the Earth's Doppler modulation. 
This includes transient-like CW signals, such as those produced by newborn NSs~\cite{Oliver:2019ksl},
or NSs in binary systems~\cite{Covas:2019jqa}.

\subsubsection{\fh{}}

\fh{} is another pipeline based on the Hough transform~\cite{Antonucci:2008jp, 2014PhRvD..90d2002A}. 
As opposed to \sh{}, binarized spectrograms are mapped onto the frequency and spindown subspace $(f_0, f_1)$. 
In this case, sky position is \emph{fixed} for all the analyzed templates.
This presents two main advantages with respect to \sh{}. 
First, it allows to increase frequency resolution independently of other parameters,
resulting in a better parameter recovery.
Second, all candidates analyzed at once are related to the same sky position. 
As a result, sky regions where templates tend to overlap with instrumental features
can be dealt with in a simpler manner~\cite{2015Leaci, 2020Intini}.
This different mapping, however, limits the generalization possibilities of \fh{}
to linear relations between parameters.

This search is usually combined with different data formats: wide parameter-space searches are usually run using
the so called Short Fourier Data Base (SFDB)~\cite{PiaSFDB}, which includes time-domain cleaning of
raw data to reduce the effect of transient noise. 
For searches at a specific parameter-space region at hand, such as supernova remnants, the Galactic center
or a specific outlier from a wider search, 
Band Sampled Data (BSD) \cite{Piccinni:2018akm, 8553107} is used to efficiently apply heterodyning filters,
reducing the computing cost of analyzing such a local parameter-space region.

Several generalizations of these methods have been developed in order to look for other kinds of source.
Power-law frequency evolution was covered using a generalization of \fh{}~\cite{Miller:2018rbg},
allowing to search for binary neutron star merger remnants~\cite{LIGOScientific:2018urg}.
A study on the suitability of \fh{} to probe planitary-mass primordial black holes 
was presented in~\cite{Miller:2020kmv}.
A method to conduct all-sky searches for CWs from the evaporation of boson clouds around spinning black holes
was developed in~\cite{DAntonio:2018sff},
using the BSD framework~\cite{Piccinni:2018akm, 8553107}.
A related method to place constraints on dark-photon dark matter interacting directly with GW interferometers was developed in~\cite{Miller:2020vsl}
and applied in \cite{LIGOScientific:2021odm}.

\subsection{Viterbi searches}

CW signals could contain stochastic contributions (spin wandering) whose behaviour
would not be well represented by the standard model introduced in Eq.~\eqref{eq:track}.
This could affect objects in both isolated and binary systems~\cite{2015PhRvD..91f2009A, Mukherjee:2017qme}, 
and should be taken into account in order to describe the underlying physics.

A simple approach is to describe the frequency-evolution model itself as a stochastic process, 
namely a Markov Chain (MC), and infer the most likely instantaneous frequency of a signal from the data.
This is usually developed under the framework of Hidden Markov Models (HMM)~\cite{Suvorova:2016rdc, 
2017PhRvD..96j2006S, 2018PhRvD..97d3013S, 2019PhRvD..99l3010S, Melatos:2021mmz}, 
but an equivalent description can be done using Bayesian probability~\cite{2019PhRvD.100b3006B}.
As discussed in~\cite{Isi:2018pzk},
the search for ultralight boson-cloud evaporation around spinning black holes could also benefit
from this kind of approach.
A specific example of such a search, directed toward Cygnus X-1 using Advanced LIGO O2 data,
was presented in~\cite{Sun:2019mqb}.

Given a set of measurements at discrete times $x = \{x_j, j=0, \dots, N-1\}$, 
we want to infer the instantaneous frequency of a signal in the data $f = \{f_j, j=0, \dots, N-1\}$.
This problem can be readily expressed as an inference one on $f$ by means of Bayes theorem~\cite{jaynes_2003}
\begin{equation}
    \mathrm{P}(f|x) \propto \mathrm{P}(f) \times \prod_{j=0}^{N-1} \mathrm{P}(x_j|f_j) 
    \label{eq:MC}
\end{equation}
where the sampling distribution of different measurements has been (conservatively) factored assuming logical
independence.
The prior probability distribution on the instantaneous frequency $\mathrm{P}(f)$ is usually specified in terms 
of the initial frequency $\mathrm{P}(f_0)$ and the transition probabilities 
of the MC $\mathrm{P}(f_j|f_{j-1})$
\begin{equation}
    \mathrm{P}(f)= \mathrm{P}(f_0)\times \prod_{j=1}^{N-1} \mathrm{P}(f_j|f_{j-1})\;.
\end{equation}
$\mathrm{P}(f_0)$ is generally taken as a uniform distribution over the searched frequency band; 
the choice of transition probabilities (thus prior probabilities) $\mathrm{P}(f_j|f_{j-1})$ 
and sampling distribution $\mathrm{P}(x_j|f_j)$ is dependent upon the method and source of interest,
as summarized in Table~\ref{tab:viterbi}.
\begin{table}
    \centering
    \begin{tabular}{cccc}
\toprule
    Method & $\ln \mathrm{P}(x_j|f_j)$ & Searches \\
\midrule
    \viterbi{} 1.0 \cite{Suvorova:2016rdc} & (Bessel-weighted) $\mathcal{F}$-statistic & \cite{LIGOScientific:2017fer, Sun:2019mqb, Millhouse:2020jlt, Beniwal:2021hvc}\\
    \viterbi{} 2.0 \cite{2017PhRvD..96j2006S} & $\mathcal{J}$-statistic & \cite{LIGOScientific:2019lyx, Middleton:2020skz, LIGOScientific:2021ozr}\\
    \viterbi{} SNR \cite{2018PhRvD..97d3013S} & $\mathcal{F}$-statistic & \cite{Jones:2020htx, LIGOScientific:2021mwx} \\
    Dual-harmonic \viterbi{} \cite{2019PhRvD..99l3010S} & $\mathcal{F}$-statistic & \cite{LIGOScientific:2021mwx} \\
    Transient \viterbi{} \cite{2019PhRvD..99l3003S} & Norm. Fourier power & \cite{LIGOScientific:2018urg, 2019PhRvD..99l3003S} \\
    \soap{} \cite{2019PhRvD.100b3006B} & Line-aware statistic & --- \\
    \viterbi{} 3.0 \cite{Melatos:2021mmz} & $\mathcal{B}$-statistic & --- \\
\bottomrule
\end{tabular}

    \caption{Summary of CW searches based on a MC implemented via the Viterbi algorithm. 
    Transition probabilities (second column) and other details of each specific pipeline are discussed in the text.
    }
    \label{tab:viterbi}
\end{table}
Regardless of the astrophysical scope, searches using a MC evolution model [Eq.~\eqref{eq:MC}] 
obtain the most likely (maximum-posterior) frequency-evolution path $f^{*}$ using the 
Viterbi algorithm~\cite{viterbi}
\begin{equation}
    f^{*} = \argmax_{f} \mathrm{P}(f|x)\;.
\end{equation}

Two basic choices exist for transition probabilities~\cite{2019PhRvD..99l3010S}, 
depending on whether the dominant frequency-drift time-scale is given by the source's spin wandering or
secular spindown. 
The former allows transitions to any neighbouring frequency bin at each time step.
The latter uses a so-called biased HMM: given a frequency at bin $j$, 
the following frequency bin must be equal or lower, $f_{j+1} \leq f_{j}$; 
the specific number of bins is usually between two and three~\cite{Jones:2020htx, Beniwal:2021hvc}.

Since no template bank is involved, \viterbi{} searches are able to benefit from a wide variety of
detection statistics at a small tuning cost. Specifically, binary modulations can be easily folded in
using the $\mathcal{J}$-statistic, which improves over the $\mathcal{C}$-statistic~\cite{Sammut:2013oba}
by combining frequency side-bands using complex weights. In order to track phase information, 
\viterbi{} 3.0 uses an efficient implementation of the $\mathcal{B}$-statistic~\cite{Prix:2009tq, Whelan:2013xka}
first proposed in~\cite{Dergachev:2011pd}.

Finally, the \soap{} pipeline~\cite{2019PhRvD.100b3006B} uses Bayesian spectral analysis 
to avoid relying on the $\mathcal{F}$-statistic,
looking for CW signals displaying a sinusoidal behaviour during short periods of time. 
To do so,
the sampling distribution is taken proportional to the data's Schuster periodogram~\cite{Bretthorst1988Bayesian},
adding an extra hypothesis to increase the robustness against strong monochromatic features in the data.
This pipeline assumes the CW signal frequency is contained in a single frequency bin;
as a result, the maximum recoverable spindown corresponds to that of a NS with a small ellipticity,
in an analogous manner to the latest \falcon{} searches~\cite{Dergachev:2020fli, Dergachev:2020upb, 
Dergachev:2021ujz}.

\subsection{Machine learning}

The use of machine-learning (ML) techniques in the search for CW signals follows one of two trends:
either to classify and summarize the results of a search's main stage, 
as discussed in Sec.~\ref{subsec:clustering},
or to substitute the search step itself, acting as a detection statistic. 
A recent review on ML applications to GW data analysis in general can be found in~\cite{Cuoco:2020ogp}.

A first approach is to train a classifier directly over Fourier-transformed raw time-series data
to distinguish the presence of a signal within background noise.
The specific data format is dependant upon the signals being looked for: 
for CW signals, which last for long periods of time over narrow frequency bands.
In~\cite{Dreissigacker:2019edy, Dreissigacker:2020xfr}, 
a convolutional neural network (CNN) is trained on the real and imaginary parts of 
short-time Fourier transforms.
Slight variations on this proposal were employed to look for postmerger signals.
These signals, 
compared to CW signals, last a shorter period of time over a wider frequency band~\cite{Sarin:2020gxb}.
In order to reflect the non-trivial time dependency of the signal, 
Ref.~\cite{Miller:2019jtp} took Fourier transforms using shorter time durations, 
finally using the data spectrogram as input for, again a CNN.

The second approach takes a less radical point of view. Instead of starting from raw data, 
machine learning algorithms are applied on the output of a search pipeline to construct a new 
detection statistic~\cite{Yamamoto:2020pus, 2020PhRvD.102h3024B}. 
This approach could be beneficial, as the output of search pipelines typically enhances signal 
features across the output parameter space. 
The specific format with which the output data is better represented, however, remains a point of discussion.\label{subsec:clustering}.

Further applications of ML to post-process the output of a search~\cite{2019arXiv190706917M, 
Beheshtipour:2020zhb, Beheshtipour:2020nko} will be covered in Sec.~\ref{subsec:clustering}.

\section{Post-processing strategies}
\label{sec:post_processing}
The main stage of a wide parameter-space search usually returns the loudest templates in terms
of a specific detection statistic. 
A good portion of these templates are correlated, 
either because their corresponding time-frequency evolution tracks sweep over similar data 
or because of the presence of non-Gaussianities in the data, 
producing broad parameter-space artifacts~\cite{Keitel:2015ova, 2018PhRvD..97h2002C, distromax}. 
The idea behind post-processing stages is to reduce the number of candidates
into an affordable quantity to be followed up.

Complementarily, veto strategies can be applied in order to reduce further the number of candidates
resulting from a search. 
A veto is a simple method to assess the consistency of a CW candidate with respect to a signal hypothesis.
The outcome is usually a boolean answer; as opposed to a post-processing or follow-up stage, 
the primary objective is to quickly reject an inconsistent CW candidate at a low computing cost.

The following subsections summarize common post-processing strategies employed in contemporary 
CW searches. After a brief overview of conicidence and clustering steps,
we review four families of vetoes. 
Other techniques employed in previous searches can be found in~\cite{2015Leaci}.

\subsection{Coincidences}

In a network of gravitational-wave detectors with a comparable level of sensitivity,
a CW signal is expected to produce significant candidates in the analysis of every detector's data.
Imposing a coincidence criterion, that is, focusing on common parameter-space regions highlighted
in every single dataset, reduces the false-alarm probability of the search, 
as noise fluctuations are less likely to be coincident across different detectors 
than CW signals~\cite{LIGOScientific:2014yew}. 
This approach obeys a robustness versus sensitivity trade-off,
as combining data from different detectors into a single analysis increases the sensitivity of a
search without increasing the required number of templates to be evaluated,
keeping the computing budget under control~\cite{Prix:2006wm}.

This approach has been widely employed in \fh{} and \sh{} searches. 
In both cases, a parameter-space distance, based on an Euclidean ansatz, 
is used to identify closeby candidates in each detector's results. 
Other searches, such as \tdfs{}, \powerflux{} or \falcon{}, impose coincidence criteria based on the overlap of
enhanced parameter-space regions across both datasets, rather than using a parameter-space distance. 

In particular, as discussed in Sec.~\ref{sec:searches},
\tdfs{} is the most prominent user of this strategy~\cite{LIGOScientific:2014yew}. 
As opposed to other semicoherent searches, it does not combine coherent segments into a semicoherent statistic, 
but looks for coincident candidates across multiple segments (including different detectors), 
reducing the overall false-alarm probability of the search.

\subsection{Parameter-space clustering}
\label{subsec:clustering}

Another option is to group together nearby templates according to some notion of distance. 
This process, usually referred to as \emph{clustering} in the data analysis literature, 
has been extensively employed by \powerflux{}, \eah{}, \fh{}, and \sh{} using different implementations,
both in terms of clustering strategy and parameter-space distance~\cite{ASI_O1, ASI_O1_full, EaT_O1, 
ASI_O2, Steltner:2020hfd, Covas:2020nwy, ASB_O3a}. 
Clustering is typically implemented using an unsupervised approach~\cite{Singh:2017kss, Tenorio:2020cqm}:
the parameter-space and the clustering algorithm itself act as prior information to construct meaningful
groupings of the resulting template bank.
Supervised ML approaches~\cite{2019arXiv190706917M, Beheshtipour:2020zhb, Beheshtipour:2020nko}, 
aimed at identifying specific parameter-space structures, have also been proposed.

Unsupervised approaches try to unveil structure from a data set. 
Prior information is encoded both in terms of the distance used to compare nearby candidates and
the linkage criteria. 
The resulting clusters are sieved through selection criteria which can take into account parameters
like the maximum significance in the cluster or its number of elements. 
We focus our exposition on the choice of parameter-space distance; 
a complete description of the clustering algorithms themselves is given in the corresponding references.

Initial implementations, used by \fh{} and \sh{}~\cite{ASI_O1, ASI_O1_full,
Covas:2020nwy} assumed a Euclidean parameter-space distance on the CW signal parameters 
$\lambda = \{f_0, f_1, \dots, \vec{n}, \dots\}$
\begin{equation}
    d(\lambda, \lambda_{*}) =  \sqrt{\displaystyle \sum_{i}\left( 
    \frac{\lambda^{(i)} - \lambda_{*}^{(i)}}{\delta \lambda^{(i)}}
    \right)^2}\;,
\end{equation}
pairing candidates within a certain distance threshold to form the final clusters.
The exact parameters $\lambda^{(i)}$ and parameter-space resolutions $\delta \lambda^{(i)}$ are
search-dependent.

A more informative approach, still based on a Euclidean ansatz, was proposed in~\cite{Singh:2017kss}.
In this case, clusters are classified according to topographic parameters by projecting 
detection statistics over planes, namely over the frequency-spindown plane and the ecliptic plane.
Instead of using the basic CW parameters,
distance was computed after projecting the sky position of the candidate onto the ecliptic plane;
thus allowing a greater variance around the ecliptic, 
where sky localization tends to become more uncertain~\cite{Krishnan:2004sv, Prix:2005gx}.

These distance measures are effective for local analyses, but quickly become unrealiable whenever more involved
parameter-space structures such as correlated parameters with periodic boundaries come into play,
as is the case for signals from sources in binary systems~\cite{Leaci:2015bka}.
A parameter-space distance for a generic, quasi-monochromatic CW signal was proposed in~\cite{Tenorio:2020cqm}
using the instantaneous detector-frame frequency associated to a CW template $f(t;\lambda)$. 
Concretely, the distance between two templates $\lambda$ and $\lambda_{*}$ can be defined as the average
mismatch between their corresponding detector-frame frequency tracks throughout an observing run
\begin{equation}
    d(\lambda, \lambda_{*}) \propto \frac{1}{T}\int^{T} 
    \mathrm{d}t \left|f(t; \lambda) - f(t; \lambda_{*})\right|\;.
\end{equation}
This prescription is consistent with the $\mathcal{F}$-statistic's parameter-space correlations 
and can be simplified into a discrete sum  for a faster implementation~\cite{Tenorio:2020cqm, 
ASB_O3a, Tenorio:2021sfj}.

Candidate post-processing, and clustering in specific, 
is also a suitable step in which machine learning strategies
are able to deliver an improvement of sensitivity. 
As opposed to raw data, on which CW signals are typically a subdominant contribution,
the structures produced by different features in the data on the parameter space of a search are suitable
to be classified using a supervised approach, as long as a clear classification of the features at hand
is available. To date, two approaches have been proposed. 

The first one~\cite{2019arXiv190706917M}, framed within the \tdfs{} pipeline, 
uses a convolutional network to classify different representations of the main search's output into
three possible classes, namely noisy bands containing Gaussian-like noise, narrow spectral artifacts,
and CW signals. A similar (albeit more manual) approach was reported in~\cite{EaT_O1}.

The second approach is developed as an alternative to the clustering algorithm employed in 
\eah{} searches~\cite{Beheshtipour:2020zhb, Beheshtipour:2020nko}. 
In this case, a neural network is trained to recognize signal-induced patterns on a certain projection of
the parameter space in order to identify typical structures associated to CW signals.
The training set is produced by manually identifying software-injected signals using an image
editing tool.

\subsection{Detector-consistency vetoes}

The first family of vetoes tests the consistency of a CW candidate across the network of GW detectors.
The basic implementation is formulated as follows: given a CW candidate with parameters $p$ and $N$ detectors, 
a detection statistic $\mathcal{S}(p)$ is computed using data from each detector alone 
$\{\mathcal{S}_{1}(p), \dots, \mathcal{S}_{N}(p)\}$ 
and combining the datasets from all detectors at once $\mathcal{S}_{\textrm{M}}(p)$. 
Then, a function of single-detector statistics 
$F\{\mathcal{S}_{1}(p), \dots, \mathcal{S}_{N}(p)\}$ is compared to the 
multi-detector statistics in order to decide whether the candidate behaves consistently with a CW signal or not.
The decision boundary, usually expressed in terms of the difference 
$F\{\mathcal{S}_{1}(p), \dots, \mathcal{S}_{N}(p)\} - \mathcal{S}_{\textrm{M}}(p)$, 
can be calibrated by means of a software-injection campaign.
As discussed in~\cite{Keitel:2013wga,Keitel:2014gha,Keitel:2015ova, Keitel:2016wtx}, 
the $\bsgl$ and $\bsgltl$ detection statistics are a Bayesian approach
to this implementation of the detector-consistency veto. 
In their implementation~\cite{LIGOScientific:2016ahk, EaT_O1, Steltner:2020hfd}, however, 
the information is processed during search-time, discarding inconsistent candidates at an 
earlier stage and potentially improving the detection of marginally significant signals.

An example of detector-consistency vetoes can be found in the \sh{} contribution to~\cite{ASI_O1},
where $F$ was taken to be the expected multi-detector statistic computed from the single-detector
statistics including the varying noise floors but ignoring the antenna pattern modulations. 
Another example is found in the \weave{} search~\cite{Wette:2021tbv}, where $F$ is simply the \emph{maximum}
detection statistic over all the involved detectors. 
A slight variation was employed in the \bsh{} analysis of early O3 data~\cite{ASB_O3a, Tenorio:2021sfj}, 
where the detection statistic of one of the LIGO detectors was compared against the other one. 
This was motivated due to the asymmetric behaviour of said detectors during the third observing run, 
with H1 more affected by noise disturbances than L1.

\subsection{\texorpdfstring{$\chi^2$}{Chi-squared} vetoes}

The second family of vetoes considers the behaviour of a putative CW signal within a particular dataset, namely,
whether the detection statistic accumulates throughout the observing run in a way which is more consistent 
with an instrumental artifact than an astrophysical signal.
This is the idea behind the $\chi^2$ veto, initially proposed in~\cite{2005PhRvD..71f2001A} for CBC signals 
and later implemented for CW signals in~\cite{2008CQGra..25r4014S}.
In this case, the dataset is partitioned into $p$ segments over which a detection statistic is computed 
$\{\mathcal{S}_1, \dots, \mathcal{S}_p\}$. 
These segment-wise statistics are then compared to the \emph{expected} segment-wise
statistics under the presence of a signal in Gaussian noise, usually characterized by a mean
and standard deviation $\mu$ and $\sigma$, and combined into a chi-squared discriminant
\begin{equation}
   \chi^2\{\mathcal{S}_1, \dots, \mathcal{S}_p\} 
    = \sum_{i = 1}^{p} \left(\frac{\mathcal{S}_i - \mu_i}{\sigma_i}\right)^2\;.
\end{equation}
Under the assumption of Gaussian noise, this discriminant follows a $\chi^2$ distribution with $p-1$ 
degrees of freedom; real-data applications, however,
must calibrate this test using a suitable injection campaign~\cite{ASI_O1}.
In~\cite{ASB_O3a, Tenorio:2021sfj}, extreme deviations of the segment-wise detection statistic 
were used to identify stretches of data in which the CW template showed a high degree of overlap
with an instrumental feature; this approach is equivalent to using a $\chi^2$ discriminant in the
limit of a very strong sample $\mathcal{S}_{j} - \mu_{j} \gg \sigma_{j}$.

\subsection{Vetoing narrow spectral features}

The third family of vetoes relies on detector characterization to identify frequency bands in which an
instrumental feature is present. 
CW searches integrate long periods of time looking for quasi-monochromatic signals 
concentrated around a fraction of a Hertz. 
Quite often, those narrow bands are populated by narrow spectral features (lines) due to instrumental or
environmental disturbances (defective power supplies, blinking LEDs, wind blowing, 
local fauna interacting with the detector\dots) which, under very general conditions, 
are able to mimic the effect produced by a CW signal in the detector, 
usually producing a high number of candidates in a search~\cite{Keitel:2014gfm, 2018PhRvD..97h2002C, LIGO:2021ppb}.
Catalogs listing narrow spectral features and their cause (if known) for the latest runs of the 
advanced detectors are publicly available~\cite{PublicLinesO1, PublicLinesO2, PublicLines, 
O3aId, O3aUid, O3Id, O3Uid}.
This kind of features are generally not a problem for CBC searches, 
as those signals sweep wide frequency bands in a relatively short time duration,
although noise subtraction techniques are applied in severe cases~\cite{LIGOScientific:2018kdd,Davis:2018yrz}.

A common approach, usually referred to as the (known) \emph{line veto} 
(see e.g.~\cite{ASI_O1, ASI_O1_full, ASI_O2, ASB_O3a}), is to check whether 
the frequency evolution of a CW candidate overlaps with any frequency band containing such instrumental
features, in which case the candidate is discarded. 
This requires a high degree of manual intervention, as line catalogs must be created and properly understood,
and incurs the risk of removing a genuine signal candidate due to an unfortunate line crossing at a
potentially insignificant period of the run.

The \dmoff{} veto was proposed in~\cite{Zhu:2017ujz} as a hypothesis-test version of the line veto:
it compares the significance of a CW candidate, which includes Doppler modulations due to the Earth's movement
around the SSB, versus the significance obtained after analyzing its surrounding frequency band using an 
unmodulated template bank (i.e. \emph{without} Doppler Modulation, hence \dmoff{}).
This veto was applied with great success in~\cite{EaT_O1}. 
As happens with the detector-consistency veto and the $\bsgl{}$ statistic~\cite{Keitel:2015ova},
the \dmoff{} veto can be refactored as a detection statistic in a Bayesian framework to construct a
proper line-robust statistic:
instead of testing against a single-detector artifact, the hypothetical $B_{\textrm{S/GMU}}$ statistic would
test against an ensemble of \emph{monochromatic} and \emph{unmodulated} signals in any number of 
detectors~\cite{Keitel:2014gfm}.

Alternatively, as performed in ~\cite{EaT_O1, Keitel:2019zhb, Steltner:2020hfd}, 
frequency bins in the data containing lines can be replaced by Gaussian noise drawn from 
the distribution of neighboring bins to suppress the presence of candidates,
preventing any candidate of instrumental origin from polluting the search results.
This process is generally performed on SFT data before starting a search.

Short-duration loud instrumental glitches also present a problem to CW searches, 
as they tend to degrade the noise floor across a wide frequency band. 
This sort of artifacts, however, 
are typically dealt with before starting a search using a cleaning
procedure such as gating~\cite{PiaSFDB, Gating, LIGO:2021ppb, Steltner:2021qjy}.

\subsection{Null-hypothesis vetoes}

The fourth family of vetoes are essentially a reformulation of the standard null-hypothesis test,
in which a CW candidate is deemed as uninteresting if it is consistent enough with respect to 
the background noise distribution.  
A simple proposal, usually referred to as \emph{off-sourcing}~\cite{Middleton:2020skz, Isi:2020uxj},
is to evaluate a CW candidate on nearly independent noise realizations by shifting its sky position away.
This is based on the fact that detector artifacts tend to imprint wider parameter-space regions with 
significant templates than CW signals.
Off-sourced time-frequency tracks are able to break signal-induced correlations 
while still being affected by instrumental disturbances. 
The resulting distribution is a proxy of the noise hypothesis' sampling distribution and can be used
to construct significance arguments about the CW candidate of interest.

This veto was adapted by~\cite{Wette:2021tbv} to evaluate the final surviving candidate of a search.
In their use case, however, they considered the distribution of the \emph{loudest} candidate from a CW search,
that is, considering the number of trials performed by evaluating a template bank on a data stream.
Such a distribution has been previously studied in the CW literature~\cite{Wette:2011eu,Dreissigacker:2018afk},
but it was not until recently that a method applicable to a generic CW search based on extreme value theory
was proposed~\cite{distromax}.

Steps towards a fully Bayesian treatment of loudest-candidate null-hypothesis 
vetoes were taken in~\cite{Tenorio:2021njf},
which proposed a Bayes factor to evaluate the loudest candidate of a CW search, $\bstar$,
whose noise-hypothesis component was constructed fitting a Gumbel distribution to the loudest outliers
of off-sourced template banks. 
This approach was used to develop a complete hierarchical follow-up framework, 
discussed in Sec.~\ref{subsec:multi_stage}.

\section{Follow-up}
\label{sec:follow_up}
Follow-up stages are contextualized within a hierarchical search, as discussed in the introduction.
They improve the parameter estimation of a CW candidate by imposing tighter constrains on its expected behaviour. 
This leads to the factual use of simple follow-up stages as signal-consistency veto strategies.
Base search stages construct less-sensitive detection statistics by effectually using less-constraining
CW signal models. For example, a semicoherent search, in which phase information is contained in discrete,
non-overlapping segments, is insensitive to arbitrary phase jumps between coherence segments. 
In this sense, 
the sensitivity loss with respect to a fully-coherent search is due to the increased trials factor of this
looser family of signal models~\cite{Dergachev:2019wqa}.

There are two ways in which follow-ups may be performed. 
Following the notation established in~\cite{Cutler:2005pn}, 
we refer to them as \emph{fresh data mode} (FDM) and \emph{recycling data mode} (RDM).
As their names suggest, FDM looks for a CW candidate in a new dataset containing brand new information;
RDM, on the other hand, re-analyzes the same dataset using a different method.
The typical example of FDM is to look for a CW candidate obtained in
a certain observing run using data from subsequent observing runs~\cite{EaT_O1, Papa:2020vfz, ASI_O3a}. 
Examples of RMD include, for example, 
multi-stage semicoherent or loosely-coherent searches aiming towards a fully-coherent
search in a restricted parameter space region~\cite{Dergachev:2010tm, 
Papa:2016cwb, Ashton:2018ure, Tenorio:2021njf}. 

Most CW searches are conducted focusing on a single dataset, 
usually the latest available observing run of the advanced detectors.
Hence, follow-up strategies operate on RDM (even though sometimes FDM assumptions are used for simplicity,
as they turn out more conservative and simpler to implement~\cite{Dreissigacker:2018afk, coherentsemicoherentF}).
Nevertheless, this sort of strategies could be detrimental towards detecting a CW signal,
as standard RDM operations leading to effectively longer coherence time may lose candidates affected by 
some form of unmodelled behaviour, such as glitches~\cite{Ashton:2017wui, Ashton:2018qth} or 
accretion-induced spin wandering~\cite{Mukherjee:2017qme}. 
Strict FDM must be used in order for a search to follow up a CW candidate with a consistent signal model.
Looking into previous observing runs, on the other hand, runs into a sensitivity problem, 
as marginal candidates may end up completely lost due to the lower quality of the detectors.
Examples of searches in which a FDM follow-up looked into a posterior observing run 
include~\cite{EaT_O1, Dergachev:2019pgs}.

\subsection{Single-stage follow-up}
\label{subsec:single_stage}

The simplest follow-up strategy calibrates a threshold on a different (more sensitive) detection statistic
according to some prescription and compares it to the score returned by reanalyzing the CW candidate.
As opposed to a veto, 
this approach points towards evaluating the consistency (or discrepancy) of the CW candidate with
respect to a certain population of signals.

The standard \fh{} follow-up as employed in~\cite{ASI_O1, ASI_O2} belongs to this category: 
baseline Fourier transform length is increased, 
imposing tighter constraints on the signal model and discarding short-duration candidates. 
Later searches also make use of the BSD framework~\cite{Piccinni:2018akm, 8553107}.
Similar strategies, in this case using the \emph{fully-coherent} $\mathcal{F}$-statistic, 
were proposed in~\cite{Shaltev:2013kqa,Shaltev:2014toa}.

In order to overcome the curse of dimensionality, 
\bsh{} searches~\cite{Covas:2020nwy, ASB_O3a, Tenorio:2021sfj} 
employ an MCMC-based \mbox{follow-up} implemented in
\pyfstat{} \cite{Keitel:2021xeq, Ashton:2018ure, Tenorio:2021njf}.
The multi-detector $\mathcal{F}$-statistic~\cite{Jaranowski:1998qm, Cutler:2005hc} 
is used to allow for arbitrarily long coherence times. 
In this sense, following up a CW candidate is equivalent to sampling the posterior probability distribution
of the phase-evolution parameters $\lambda$ given a data stream~\cite{Ashton:2018ure}
\begin{equation}
    \mathrm{P}(\lambda | x) \propto  e^{\mathcal{F}(\lambda;x)} \cdot \mathrm{P}(\lambda) \;,
\end{equation}
where the prior probability distribution $\mathrm{P}(\lambda)$ represents the parameter-space region of
interest identified by a search. The result of this grid-less approach is the $\mathcal{F}$-statistic
evaluated at the loudest candidate of the parameter space at a negligible mismatch.
This approach generalizes that of~\cite{Shaltev:2013kqa,Shaltev:2014toa}, 
which specializes in the single-stage fully-coherent follow-up of CW candidates.

As discussed in Sec.~\ref{subsec:multi_stage}, 
the use of MCMC methods simplifies the setup of a generic multi-stage follow-up, 
as no calibration of parameter-space grids are required. 
The onus in this case is on the search pipeline to deliver a small-enough prior support for the MCMC to converge. 
This can always be achieved by starting from a shorter coherence time~\cite{Ashton:2018ure},
at the expense of increasing the number of follow-up stages.

A similar strategy is used by the \tdfs{} follow-up, focusing on the optimization aspect of the procedure.
In this case, a max-finding algorithm such as \cite{Nelder1965ASM} is employed to travel around the parameter
space. The result, as with \texttt{PyFstat}, are the most favored signal parameters, 
corresponding to the ones reporting the loudest $\F$-statistic value. 

\subsection{Multi-stage follow-up}
\label{subsec:multi_stage}

As discussed during Sec.~\ref{sec:introduction}, 
multi-stage follow-ups are the natural continuation to a wide parameter-space search after identifying
interesting parameter-space regions. 
Given a data stream, each subsequent follow-up stage operates in RDM, 
gradually increasing the coherence time with respect to previous stages and, as a consequence, 
imposing a tighter version of the selected signal model. 
The effect on a CW candidate is twofold: first, non-astrophysical CW-like artifacts tend to get rejected
as coherence time increases (though, as earlier discussed, 
this could have detrimental effects on more complex CW signals too);
second, increasing coherence time results in a refinement in parameter-space resolution, 
improving the parameter estimation of the candidate at hand. 
This last effect must be considered carefully, 
as it also implies an increased number of templates to analyze, 
quickly becoming unaffordable if parameter-space regions are not gradually narrowed down.

We start by discussing the follow-up strategy introduced in~\cite{Papa:2016cwb}, 
which was applied as a generic follow-up to multiple CW searches with minor 
modifications~\cite{ASI_O1, ASI_O1_full, EaT_O1, Steltner:2020hfd}.
This example is paradigmatic in the sense that it fully exposes the two main challenges of a multi-stage setup.
A similar approach, albeit at much smaller scale, was employed to follow-up \sh{} results in~\cite{ASI_O2}.
First, one must set up a proper set of parameter-space grids such that CW signals are not lost due to a bad
parameter-space covering. If an analytical model of the follow-up method at hand is not available, 
as is the case in ~\cite{Papa:2016cwb}, 
one must resort to an extensive software injection campaign to construct a suitable setup. 
Second, a criterion must be set up to select/reject CW candidates after performing different stages. 
If the detection statistic's behaviour across different stages is well understood 
(see e.g.~\cite{coherentsemicoherentF}), an analytical criterion can be derived from first principles;
otherwise, a software injection campaign must be used to calibrate a rejection criterion.

Latest developments on the follow-up of CW candidates~\cite{Ashton:2018ure, Tenorio:2021njf} 
are able to simplify the setup for the $\F$-statistic,
although further work is required for its application to generic detection statistics.
As discussed in Sec.~\ref{subsec:single_stage}, 
the use of MCMC samplers simplifies the setup due to their lack of grids: if a CW signal is within the 
prior support, parameter-space samplers can get arbitrarily close to the injection parameters given enough
time to wander around the prior volume. This argument can be posed quantitatively in terms of the so-called
coherence time \emph{ladder}~\cite{Ashton:2018ure}, 
which make use of the parameter-space metric to increase the parameter-space resolution in a controlled manner.
Since the follow-up is typically a local analysis, 
rough estimates of the number of templates using typical parameter-space resolutions are usually a valid
approximation~\cite{Ashton:2018qth}.

Proper comparison of detection statistics from different stages in a Bayesian framework 
is currently restricted to the $\mathcal{F}$-statistic~\cite{coherentsemicoherentF},
as the sampling distribution under a signal hypothesis must be known 
given a (squared) signal-to-noise ratio $\rho^2$.
This result was used in~\cite{Tenorio:2021njf} to propose $\mathcal{B}^{*}_{\textrm{S}/\textrm{N}}$,
a (meta) Bayes factor (as the $\mathcal{F}$-statistic itself is a Bayes factor)
evaluating the result of a multi-stage follow-up, 
pushing forward the development of a fully-Bayesian follow-up of CW candidates. 
In this case, the probability under the noise hypothesis was derived from a combination of 
off-sourcing and extreme value theory~\cite{distromax}. 
The use of detection statistics (Bayes factors) as data proxies to construct Bayesian arguments
is also discussed in~\cite{Rover:2011zq, Dreissigacker:2018afk}.

An alternative family of follow-up methods were developed under the name of \emph{loose coherence}
\cite{Dergachev:2010tm, Dergachev:2011pd, Dergachev:2018ftg}, already introduced in Sec.~\ref{sec:falcon}. 
In this case, instead of following the ad hoc recipe of increasing coherence time until a fully-coherent
search is achieved, phase information across neighbouring time segments is gradually correlated 
in a controlled manner by combining (complex) Fourier amplitudes. 
As opposed to semicoherent methods, which allow for arbitrary phase jumps at the border of a segment,
loosely coherent methods allow for phase shifts within pre-specified ranges, 
depending on the required robustness of the method.
Under this framework, semicoherent methods are rediscovered imposing delta-correlation between phases 
at consecutive time segments. This follow-up approach is fully integrated within the \powerflux{} and \falcon{}
searches~\cite{ASI_O1, ASI_O1_full, Dergachev:2019pgs, Dergachev:2019wqa, Dergachev:2019oyu, Dergachev:2020fli,
Dergachev:2020upb}

\section{Upper bounds on \texorpdfstring{$h_0$}{CW amplitude}}
\label{sec:sensitivity}
Since no CW detection has been reported to date, 
the main data product of CW searches are bounds on the nominal gravitational-wave amplitude $h_0$ produced
by a population of sources consistent with the target signal model.
As discussed in Sec.~\ref{sec:introduction}, 
astrophysical information can be extracted from this quantity by taking different assumptions,
such as the maximum allowed ellipticity from a galactic neutron star at a certain distance from the detector.

Two basic approaches are pursued at production level to derive said upper bounds, 
depending on whether the aim is for a strict frequentist upper limit or population-based sensitivity estimations. 
The calibration and establishment of upper bounds of any kind usually involves an extensive software injection
campaign in real data with a non-negligible computing cost; due to this, a common approach lies in between
both extrema, quoting a proper estimation at a definite set of representative frequency bands and interpolating
the results across the rest of the spectrum.

Population-based sensitivity estimations are based on estimating the false-dismissal probability of a search
given a certain setup (be it a threshold at a fixed false-alarm probability or a more intricate procedure).
The $p\%$ detection probability amplitude $h_0^{p\%}$ corresponds then to the amplitude $h_0$ associated to
a false dismissal of $(1-p)\%$ after properly marginalizing with respect to other amplitude parameters
using a set of priors reflecting the studied source distribution~\cite{Wette:2011eu}. 
In practise~\cite{ASI_O1, ASI_O1_full, EaT_O1, ASI_O2, Steltner:2020hfd, ASB_O3a, Covas:2020nwy},
detection probabilities are usually estimated numerically by means of an injection recovery campaign.

Strict frequentist upper limits, on the other hand, 
return a conservative estimate of the upper bound, 
in the sense that false-dismissal probability is~\emph{at most} $(1-p)\%$. 
An example of this kind is the universal statistic procedure~\cite{Dergachev:2012jm}, 
which is able to construct strict frequentist upper limits regardless of the underlying noise distribution.
This procedure has been extensively combined with the \powerflux{} and \falcon{} pipelines to efficiently
produce robust upper limits under different GW polarization assumptions~\cite{ASI_O1, 
ASI_O1_full, Dergachev:2019pgs, Dergachev:2019wqa, Dergachev:2019oyu, Dergachev:2020fli, Dergachev:2020upb}.

To date, all wide parameter-space searches have made use of one of these two upper bounds to report on
their results. These upper bounds, however, describe the probability of detecting a signal given an 
ensemble of equivalent noise realizations, 
rather than the range where a signal could be found given the data stream at hand.
Work towards reporting the latter, Bayesian upper bounds, for wide parameter-space searches 
was developed in~\cite{Rover:2011zq, Dreissigacker:2018afk}.

\section{Summary}
\label{sec:summary}
We reviewed the methods employed by current wide parameter-space searches 
for continuous gravitational waves from unknown sources conducted on advanced-detector data. 
The most widespread approach consists of a hierarchical setup in which parameter-space regions are 
analyzed using more sensitive (and consequently more expensive) methods as they are gradually narrowed-down.
Detecting a CW signal requires both an instrumental and computational effort
to confidently unveil such a weak signal using the current generation of gravitational-wave detectors.
The use of multiple methods taking different tradeoffs in sensitivity and robustness against
instrumental artifacts provides an ideal environment to pursue new strategies towards CW detection
and parameter estimation. 
For the sake of completeness, Table~\ref{table:summary} provides a comprehensive summary of the search
methods reviewed during the present work.

\begin{table}
    \centering
\begin{tabular}{rlc}
    \toprule
        Search & Pipeline & References \\
    \midrule
        All-sky O1 
           & \eah{} & \cite{EaT_O1} \\ 
           & \falcon{} & \cite{Dergachev:2019wqa, Dergachev:2019oyu}\\
           & \fh{} & \cite{ASI_O1} \\
           & \powerflux{} & \cite{ASI_O1, ASI_O1_full} \\
           & \sh{} & \cite{ASI_O1, ASI_O1_full} \\
           & \tdfs{} & \cite{ASI_O1, ASI_O1_full} \\
        All-sky O2 
           & \bsh{} & \cite{Covas:2020nwy}\\
           & \eah{} & \cite{Steltner:2020hfd}\\
           & \falcon{} & \cite{Dergachev:2020fli,Dergachev:2020upb, Dergachev:2021ujz} \\
           & \fh{} & \cite{ASI_O2} \\
           & \sh{} &  \cite{ASI_O2} \\
           & \tdfs{}  &  \cite{ASI_O2} \\
        All-sky O3a 
            & \bsh{}  & \cite{ASB_O3a} \\
            & \powerflux{} & \cite{ASI_O3a} \\
    \midrule
        Deep exploration O2 & \weave{} & \cite{Wette:2021tbv} \\
    \midrule
        GC O1 & \powerflux{} & \cite{Dergachev:2019pgs}\\
        GC O2 & \fh{} + BSD & \cite{Piccinni:2019zub} \\
    \midrule
        SNR O1 
             & \eah{} & \cite{Ming:2019xse} \\
             & Fully-coherent $\F$-statistic & \cite{LIGOScientific:2018esg} \\
        SNR O2 
             & \eah{} & \cite{Papa:2020vfz, Ming:2021xtz} \\
             & Fully-coherent $\F$-statistic & \cite{Lindblom:2020rug} \\
             & \viterbi{} 1.0 & \cite{Millhouse:2020jlt} \\
             & \viterbi{} SNR & \cite{Jones:2020htx} \\
        SNR O3a 
              & \fh{} + BSD & \cite{LIGOScientific:2021mwx}\\
              & Dual-harmonic \viterbi{} & \cite{LIGOScientific:2021mwx}\\
              & \viterbi{} SNR & \cite{LIGOScientific:2021mwx}\\
    \midrule
        CDOs in the Solar System O2 & Excess power & \cite{Horowitz:2019pru} \\
    \midrule
        Cygnus X-1 O2 & \viterbi{} 1.0 & \cite{Sun:2019mqb} \\
    \midrule
        Scorpius X-1 O1 
            & \crosscorrelation{} & \cite{LIGOScientific:2017zez} \\
            & \viterbi{} 1.0 & \cite{LIGOScientific:2017fer} \\
        Scorpius X-1 O2  
            & \crosscorrelation{} & \cite{Zhang:2020rph}\\
            & \viterbi{} 2.0 & \cite{LIGOScientific:2019lyx} \\
    \midrule
        LMXBs O2 & \viterbi{} 2.0 & \cite{Middleton:2020skz} \\
    \bottomrule
\end{tabular}

    \caption{Summary of CW search methods covered by the present review, grouped by scope and observing run.} 
    \label{table:summary}
\end{table}

\vspace{6pt} 
\newpage
\authorcontributions{All authors have read and agreed to the published version of the manuscript.}
\funding{
    This work was supported by European Union FEDER funds, 
    the Spanish Ministerio de Ciencia e Innovación,
    and the Spanish Agencia Estatal de Investigaci\'on grants
    PID2019-106416GB-I00/AEI/MCIN/10.13039/501100011033,  
    RED2018-102661-T,    
    RED2018-102573-E,    
    Comunitat Aut\`onoma de les Illes Balears through the Conselleria de Fons Europeus, Universitat i Cultura
    and the Direcci\'o General de Pol\'itica Universitaria i Recerca with funds from the Tourist Stay Tax Law ITS 2017-006 (PRD2018/24, PRD2020/11),
    Generalitat Valenciana (PROMETEO/2019/071),  
    EU COST Actions CA18108, CA17137, CA16214, and CA16104.
    R.~T.~is supported by the Spanish Ministerio de Universidades (ref.~FPU 18/00694).
    D.~K.~is supported by the Spanish Ministerio de Ciencia, Innovación y Universidades (ref.~BEAGAL 18/00148)
    and cofinanced by the Universitat de les Illes Balears.
}

\acknowledgments{
    We thank Karl Wette, Keith Riles, Ornella J. Piccinni,
    and the LIGO-Virgo-KAGRA Continuous Wave working group for useful suggestions.
    This paper has been assigned document number LIGO-P2100408.
}

\conflictsofinterest{The authors declare no conflict of interest.}

\reftitle{References}


\externalbibliography{yes}
\bibliography{references.bib}

\end{paracol}
\end{document}